\documentclass[twocolumn,trackchanges]{aastex701}

\usepackage{amsmath}
\usepackage{natbib}

\usepackage{float}
\usepackage{multirow}
\usepackage{array}
\usepackage{booktabs}

\begin{document}

\title{Cross-validation of six dispersion measure estimation methods for FRB~20240114A}

\author[orcid=0009-0008-4576-3120]{Tong-Lun Wang}
\affiliation{School of Physics, Huazhong University of Science and Technology, Wuhan 430074, China}
\affiliation{Purple Mountain Observatory, Chinese Academy of Sciences, Nanjing 210023, China}
\email{m202470280@hust.edu.cn}  

\author[orcid=0000-0003-2366-219X]{Song-Bo Zhang*}
\affiliation{Purple Mountain Observatory, Chinese Academy of Sciences, Nanjing 210023, China}
\affiliation{CSIRO Space and Astronomy, Australia Telescope National Facility, PO Box 76, Epping, NSW 1710, Australia}
\email[show]{sbzhang@pmo.ac.cn}

\author[0000-0002-5400-3261]{Yuan-Chuan Zou*} 
\affiliation{School of Physics, Huazhong University of Science and Technology, Wuhan 430074, China}
\email[show]{zouyc@hust.edu.cn}

\author[0009-0009-3517-6640]{Xuan Yang}
\affiliation{Purple Mountain Observatory, Chinese Academy of Sciences, Nanjing 210023, China}
\affiliation{School of Astronomy and Space Sciences, University of Science and Technology of China, Hefei 230026, China}
\email{yangxuan@pmo.ac.cn}

\author[0000-0002-3386-7159]{Pei Wang*}
\affiliation{State Key Laboratory of Radio Astronomy and Technology, National Astronomical Observatories, Chinese Academy of Sciences, Beijing 100101, China}
\affiliation{Institute for Frontiers in Astronomy and Astrophysics, Beijing Normal University, Beijing 102206, China}
\email[show]{wangpei@nao.cas.cn}

\author[orcid=0000-0002-4304-2759]{Di Xiao}
\affiliation{Purple Mountain Observatory, Chinese Academy of Sciences, Nanjing 210023, China}
\affiliation{State Key Laboratory of Radio Astronomy and Technology, Purple Mountain Observatory, Chinese Academy of Sciences, 10 Yuanhua Road, Nanjing 210023, China}
\email{dxiao@pmo.ac.cn}

\author[orcid=0000-0002-6165-0977]{Xiang-Han Cui}
\affiliation{State Key Laboratory of Radio Astronomy and Technology, National Astronomical Observatories, Chinese Academy of Sciences, Beijing 100101, China}
\affiliation{ASTRON, the Netherlands Institute for Radio Astronomy, Oude Hoogeveensedijk 4,7991 PD Dwingeloo, The Netherlands}
\email{cuixianghan@nao.cas.cn}

\author[orcid=0000-0001-5931-2381]{Ye Li}
\affiliation{Purple Mountain Observatory, Chinese Academy of Sciences, Nanjing 210023, China}
\email{yeli@pmo.ac.cn}

\author[orcid=0000-0002-9586-7904]{Hao Qiu}
\affiliation{SKA Observatory, Bentley, WA 6102, Australia}
\email{hao.qiu@skao.int}

\author[0009-0004-5530-4447]{Ying-Ze Shan}
\affiliation{School of Physics, Huazhong University of Science and Technology, Wuhan 430074, China}
\email{shanyingze@hust.edu.cn}

\author[0000-0002-7949-3906]{Jun-Yi Shen}
\affiliation{School of Physics, Huazhong University of Science and Technology, Wuhan 430074, China}
\email{D202480075@hust.edu.cn}

\author[orcid=0009-0002-7318-435X]{Ya Zeng}
\affiliation{School of Physics, Huazhong University of Science and Technology, Wuhan 430074, China}
\email{2739706499@qq.com}

\author[orcid=0009-0002-3020-9123]{Long-Xuan Zhang}
\affiliation{School of Physics, Huazhong University of Science and Technology, Wuhan 430074, China}
\email{d202580060@hust.edu.cn}

\author[orcid=0009-0008-6247-0645]{Wen-Long Zhang}
\affiliation{Purple Mountain Observatory, Chinese Academy of Sciences, Nanjing 210023, China}
\affiliation{School of Astronomy and Space Sciences, University of Science and Technology of China, Hefei 230026, China}
\email{wlzhang@pmo.ac.cn}

\begin{abstract}
Fast Radio Bursts (FRBs) are important cosmological probes, but their applications depend critically on accurate dispersion measure (DM) determinations. We present a systematic comparison of six DM estimation methods using 2,874 bursts from FRB~20240114A, the most active repeating FRB currently known, observed by FAST during a single 4.4-hr session on 2024 March 12. This large, homogeneous sample over a short timescale, during which the propagation environment is expected to be nearly static, provides an ideal benchmark for isolating algorithmic effects on DM determination. 
We investigate the dependence of inter-method consistency on signal-to-noise ratio (S/N), burst morphology, and radio frequency interference (RFI). Low-S/N bursts exhibit significantly larger inter-method deviations, while single-component bursts produce highly consistent DM values across methods. In contrast, complex double- and multiple-component bursts with drifting substructures lead to substantial inter-method scattering, indicating that DM discrepancies are primarily driven by algorithmic responses to burst morphology. RFI does not significantly alter the global statistical behavior of DM deviations, but it affects density-filtering methods through morphology distortion caused by frequency-channel masking. Even after imposing strict inter-method consistency constraints, FRB\,20240114A still exhibits notable apparent DM fluctuations spanning $\sim$528--534~pc~cm$^{-3}$ over 15,780~s. For morphologically simple bursts these variations far exceed the measurement uncertainty and, on second-to-minute timescales, cannot arise from any plausible change in the line-of-sight electron column, pointing instead to a frequency-dependent emission-time structure intrinsic to the bursts that mimics dispersion. These results highlight the necessity of a reliable FRB DM measurements and provide a reference for future high-precision FRB studies.

\end{abstract}



\section{Introduction}

Fast Radio Bursts (FRBs) are millisecond-duration radio transients with luminosities exceeding $10^{36}$ erg s$^{-1}$. Their dispersion measures (DMs) far exceed the contribution from the Galactic interstellar medium, implying an extragalactic and likely cosmological origin. Since the discovery of the first FRB by \citet{1}, the known population has grown rapidly \citep{2}. A breakthrough came with the discovery of the first repeating source, FRB 20121102A, by \citet{3}. Interferometric observations with the VLA and VLBI precisely localized this source \citep{4,5} to a low-metallicity dwarf galaxy at $z\sim0.19$ \citep{6}, firmly establishing the cosmological origin of FRBs and demonstrating that some sources repeat.

New advanced facilities including FAST \citep{fast1,fast2,fast3,fast4},  ASKAP \citep{askap1,askap2,askap3,askap4} and CHIME \citep{chime1,chime2,chime3,chime4} have dramatically increased the detection rate and enabled detailed studies of burst time--frequency structures \citep{51,51.5}. Magnetar-based models are the leading explanation for FRB production \citep{7}, particularly after the detection of an FRB-like burst from the Galactic magnetar SGR 1935+2154 accompanied by hard X-rays \citep{8,9}. Alternative scenarios involving compact binaries have also been proposed \citep{10}.

The DM characterizes the integrated column density of free electrons along the line of sight, with contributions from the Galactic ISM, Galactic halo, intergalactic medium (IGM), and host galaxy. Most FRBs show DMs significantly exceeding the Galactic predictions from the NE2001, NE2025 and YMW16 models \citep{NE,ne2025,YMW}. The IGM contribution follows an approximately linear DM--$z$ relation (the Macquart relation), which provides strong evidence for the cosmological origin of FRBs and establishes DM as a key cosmological observable.

The cosmological applications of FRB DMs include constraining the dark energy equation of state, measuring $H_0$ \citep{Liu2026}, and addressing the missing baryon problem. \citet{Deng2014} derived the cosmological DM framework, demonstrating that FRBs can probe baryon distribution, dark energy, and the ionization history of the Universe. \citet{McQuinn2014} showed that scatter in the DM--$z$ relation reflects baryon inhomogeneity in the cosmic web and can constrain feedback processes in galactic halos. Observationally, \citet{41} confirmed the DM--$z$ relation with a sample of precisely localized FRBs, providing direct evidence that a substantial fraction of missing baryons reside in diffuse IGM gas.

These developments have been extended further \citep{129,128}. By combining FRB observations with large-scale structure surveys, DM differences between sightlines intersecting and avoiding cosmic filaments can directly measure baryon overdensity within filamentary structures \citep{41.5}. The differential DM--$z$ relation enables direct $H(z)$ measurements at different redshifts, requiring only several hundred FRBs with redshift information for competitive cosmological constraints \citep{46}. DM observations can also constrain cosmic reionization history, particularly helium reionization \citep{Zheng2014,42}.

Beyond cosmology, DM is crucial for FRB signal searching and data processing. Accurate DM measurements recover fine temporal structures and improve signal-to-noise ratio (S/N), enhancing the detectability of weak bursts \citep{2003Searches}. Traditional DM estimation maximizes S/N after de-dispersion, which is effective for simple morphologies but can fail for complex bursts. Optimized methods include autocorrelation-based approaches \citep{52}, forward derivative maximization \citet{51}, and power-spectrum analyses using tools such as \texttt{DM\_phase} \citep{54} and \texttt{DM\_power} \citep{55}. However, systematic discrepancies among methods can propagate into redshift estimation, energy function determination, population statistics, and cosmological parameter inference. A systematic investigation of DM estimation methods and their accuracy is therefore essential for understanding FRB propagation physics, source environments, and the reliability of cosmological applications.

Given the importance of DM precision and the discrepancies among existing techniques for complex burst morphologies, we present a systematic comparison of six DM measurement methods using 2,874 bursts from FRB 20240114A observed on 2024 March 12. We investigate the relationship between burst morphology and algorithmic response, quantify the dominant sources of systematic uncertainty, and evaluate method consistency under different observational conditions. The methods are described in Section~\ref{Method}, the data in Section~\ref{Data}, results in Section~\ref{Results}, discussion in Section~\ref{Discussion}, and conclusions in Section~\ref{Conclusion}.

\section{Method}
\label{Method}
\subsection{Dispersion Measure}
The DM is defined as the integrated column density of free electrons along the line of sight \citep{lorimer2005handbook}:
\begin{equation}
    {\rm DM}=\int_{0}^{d} n_e(l)\,dl,
\end{equation}
where $n_e$ is the electron number density and $d$ is the propagation distance. 

As radio waves propagate through an ionized plasma, lower-frequency signals experience a greater dispersive delay:
\begin{equation}
    \Delta t=\frac{e^2}{2\pi m_e c}\left(\nu^{-2}-\nu_0^{-2}\right){\rm DM},
\end{equation}
where $e$ and $m_e$ are the electron charge and mass, $c$ is the speed of light, and $\nu$ and $\nu_0$ are the observing and reference frequencies.

\subsection{${\rm DM}$ Estimation Methods}
Before de-dispersion, FRB dynamic spectra show a frequency-dependent sweep from dispersive delays. DM determination involves two steps: de-dispersing the data over a grid of trial DMs, then evaluating the de-dispersed signal with an optimization metric.

There is no exact analytical expression for DM \citep{DM_k}. DM values are estimates derived from the dispersion slope using different algorithms. Given the diverse burst morphologies, multi-method cross-validation is essential. We employ six methods:
\begin{enumerate}
    \item Peak flux maximization
    \item S/N maximization
    \item Forward derivative maximization
    \item Power spectrum optimization
    \item Kurtosis maximization with Density filtering
    \item Entropy minimization with Density filtering
\end{enumerate}

Methods 1 and 2 are signal-optimization approaches: they maximize pulse intensity and S/N by adjusting DM. Methods 3 and 4 are structure-optimization approaches: they enhance the intrinsic time-frequency coherence of the burst. Methods 5 and 6 are newly proposed statistical approaches: using density filtering with kurtosis and entropy analyses, they measure DM from the distribution and concentration of signal in signal density space (SDS).

\subsubsection{Peak flux maximization}
The optimal DM is determined by maximizing the peak flux of the frequency-integrated time series. For each trial DM, the dynamic spectrum is summed along the frequency axis, and the peak flux is
\begin{equation}
{\rm F_{peak}} = \frac{I_{\rm peak}}{\sigma},
\end{equation}
where $I_{\rm peak}$ is the peak intensity and $\sigma$ is the off-pulse noise standard deviation.

\subsubsection{S/N maximization}
Boxcar filters of different widths are applied to the de-dispersed time series to enhance signals with varying durations. For a boxcar width $W$:
\begin{equation}
{\rm S/N_{boxcar}} = \frac{1}{\sqrt{N_{\rm box}}\sigma}\sum_{i=1}^{W}I_i,
\end{equation}
where $I_i$ is the signal intensity at sample $i$, $\sigma$ is the noise standard deviation, and $N_{\rm box}$ is the number of samples in the boxcar. The maximum S/N over all boxcar widths is selected at each trial DM, and the global S/N maximum determines the optimal DM.

\subsubsection{Forward derivative maximization}
Following \citet{51}, the sharpness of the de-dispersed burst profile is characterized by the forward difference operator:
\begin{equation}
D=\lim_{\delta t \to 0}\frac{I(t+\delta t)-I(t)}{\delta t}
\approx
\frac{I(t+dt)-I(t)}{dt},
\end{equation}
where $dt$ is the temporal resolution and $I(t)$ is the signal intensity at time $t$.

A correctly de-dispersed burst produces the sharpest temporal structure and the largest forward derivative. The optimal DM maximizes this metric over the trial DM range.

\subsubsection{Power spectrum optimization}
This method estimates DM by evaluating the coherence of the de-dispersed signal in Fourier space. The dominant implementations are \texttt{DM\_power} \citep{55} and \texttt{DM\_phase} \citep{54}. Our algorithm builds upon the \texttt{DM\_power} framework. Unlike \texttt{DM\_power}, which refines within a narrow range around a given reference DM, our implementation performs a wide-range blind search and outputs DM directly.

\subsubsection{Density filtering}
Density filtering applies an intensity threshold to the dynamic spectrum, transforming the burst into SDS \citep{2025Detecting}:
\begin{equation}
S'(t,\nu)=
\begin{cases}
1, & S(t,\nu)\geq T \\
0, & S(t,\nu)<T
\end{cases},
\end{equation}
where $S(t,\nu)$ is the burst intensity at time $t$ and frequency $\nu$, and $T$ is a predefined threshold. This binary transformation suppresses noise while preserving significant signal structures.

From the SDS representation, we construct two statistical optimization criteria:

\paragraph{Kurtosis maximization}
The frequency-integrated SDS profile is obtained by summing along the frequency axis:
\begin{equation}
I(t)=\sum_{\nu}S'(t,\nu).
\end{equation}
The kurtosis of the integrated profile is defined as
\begin{equation}
K=\frac{\langle(I-\mu)^4\rangle}{\sigma^4},
\end{equation}
where $\mu$ and $\sigma$ are the mean and standard deviation of $I(t)$, respectively. A more compact burst profile gives a larger kurtosis. The optimal DM maximizes $K$.

\paragraph{Entropy minimization}
The SDS entropy is defined as
\begin{equation}
H=-\sum_t p(t)\log p(t),
\end{equation}
where
\begin{equation}
p(t)=\frac{I(t)}{\sum I(t)}
\end{equation}
is the normalized intensity distribution. Lower entropy corresponds to a more compact, coherent burst. The optimal DM minimizes $H$.

\subsection{Uncertainty Estimation}
This uncertainty definition is also applied to the peak flux maximization method.
Several schemes exist for S/N maximization uncertainties \citep{53,qiu2023}. We define the uncertainty as the DM range corresponding to a 5\% drop from the peak S/N, applied to both peak flux and S/N maximization.

For the Power spectrum optimization, uncertainties are estimated using bootstrap resampling \citep{55}. For forward derivative maximization, kurtosis maximization, and entropy minimization methods, uncertainties are derived from the $1\sigma$ confidence interval of the Gaussian centroid, obtained by least-squares fitting of the metric–DM relation and propagated via the parameter covariance matrix.

Finite instrumental resolution also introduces systematic uncertainty \citep{qiu2023}. We incorporate both temporal and frequency resolution effects.

The total timing uncertainty is
\begin{equation}
\Delta t=
\sqrt{
\delta t^2+\delta t_{\nu}^2
},
\end{equation}
where $\delta t$ is the intrinsic time resolution and $\delta t_{\nu}$ is the intra-channel dispersive smearing:
\begin{equation}
\delta t_{\nu}=K\left[\frac{1}{(\nu_c-\delta\nu)^2}-\frac{1}{(\nu_c+\delta\nu)^2}\right]{\rm DM},
\end{equation}
with
\begin{equation}
K=\frac{e^2}{2\pi m_e c},
\label{eq:K}
\end{equation}
where $\nu_c$ is the channel centre frequency, and $\delta\nu$ is half of the channel bandwidth.

The DM uncertainty from timing error is
\begin{equation}
\delta \mathrm{DM_{res}}
=
\frac{
\Delta t
}{
K\left(
\frac{1}{\nu_{\rm low}^2}
-
\frac{1}{\nu_{\rm high}^2}\right)
},
\end{equation}
where $\nu_{\rm low}$ and $\nu_{\rm high}$ are the lower and upper band edges.

Combining algorithmic and resolution uncertainties:
\begin{equation}
\sigma_{\rm DM} =\sqrt{\left ( \delta {\rm DM}_{met} \right )^{2}+\left ( \delta {\rm DM}_{res} \right )^{2}  } 
\end{equation}
where $\delta {\rm DM}_{met}$ denotes the uncertainty arising from the measurement method.
We assume resolution uncertainty is independent of methodological uncertainty.

\subsection{Inter-method Scatter}
We define the standard deviation of the six DM estimates for each burst as
\begin{equation}
    \mathrm{std}_i = \sqrt{\frac{1}{n-1}\sum_{k=1}^{n}\left({\rm DM}_{i,k} - \overline{\rm DM}_{i}\right)^2},
    \label{eq:std_i}
\end{equation}
where std$_i$ characterizes the deviation of DM values for the $i$-th sample, $n=6$ is the total number of measurement methods, $k$ represents the DM measured by the $k$-th method, and $\overline{\rm DM}$ is the mean DM across all methods. Small std$_i$ indicates strong inter-method agreement.

Furthermore, to characterize measurement discrepancies between pairwise methods, we introduce the statistic $\mathrm{std}_{kl}$, which is defined as the standard deviation of DM differences between the $k$-th and $l$-th estimators over all samples:
\begin{equation}
    \mathrm{std}_{kl} = \sqrt{\frac{1}{N-1}\sum_{i=1}^{N}\left[({\rm DM}_{i,k} - {\rm DM}_{i,l}) - \overline{\Delta {\rm DM}_{kl}}\right]^2}
    \label{eq:std_nl}
\end{equation}
\noindent where $N$ is the number of bursts in the subsample, ${\rm DM}_{i,k}$ denotes the DM value of the $i$-th burst estimated using the $k$-th method, and $\overline{\Delta {\rm DM}_{kl}} = \frac{1}{N}\sum_{i=1}^{N}({\rm DM}_{i,k} - {\rm DM}_{i,l})$ represents the mean DM difference between the two estimators.

\section{Data}
\label{Data}

A robust assessment of DM measurement accuracy requires a large sample of bursts within a short time window, so that the true DM and propagation environment are unlikely to vary significantly. FRB\,20240114A is currently the most active known repeating FRB \citep{FRB20240114A_FAST,S_N1}. On 2024 March 12, FAST detected over 3,200 bursts from this source in a single 4.4-hr observation \citep{zhang2026investigating}, providing the largest homogeneous burst sample available for DM method comparison. Over this duration, the source's local plasma environment and the line-of-sight ionized column are expected to remain nearly constant, making this dataset an ideal benchmark for isolating algorithmic effects from true astrophysical DM variations.

We use the public dataset from \citet{zhang2026investigating}, covering 1.0--1.5~GHz with a native time resolution of 49.152~$\mu$s. After discarding incomplete and truncated bursts, we retain 2,874 events. The data have been preprocessed with burst searching and RFI mitigation but remain non-de-dispersed. The mean DM is approximately $529.2~{\rm pc~cm^{-3}}$.

We apply all six methods to these 2,874 bursts. The DM search range is $515$--$545~{\rm pc~cm^{-3}}$ with a step of $0.1~{\rm pc~cm^{-3}}$.

\section{Results}
\label{Results}

\begin{figure}
	\centering
	\includegraphics[width=0.95\columnwidth]{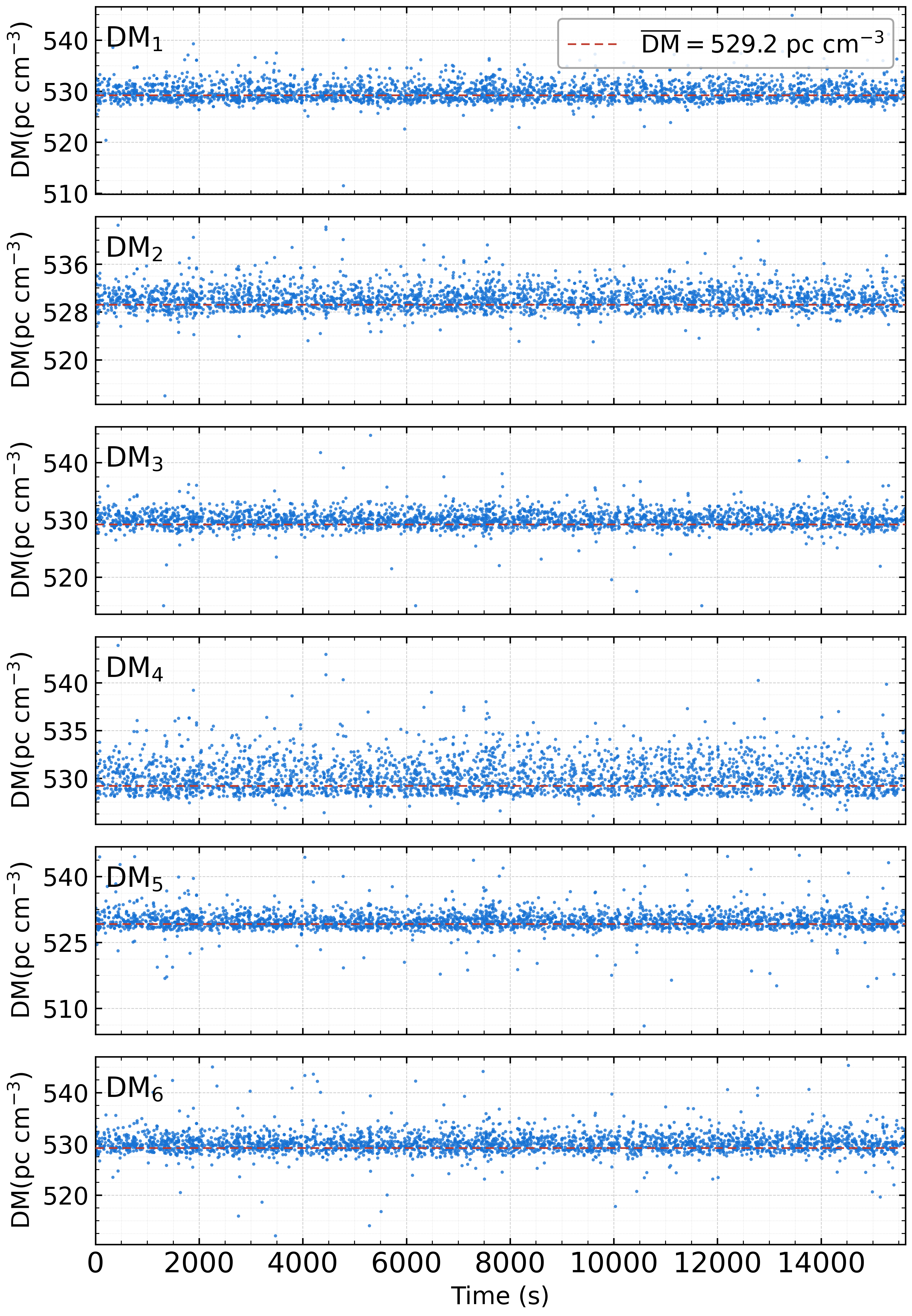}
	\caption{Temporal variations of DM for 2,874 bursts of FRB 20240114A observed on March 12, 2024. Each panel shows the DM time series obtained by one of the six measurement methods (labeled ${\rm DM}_1$--${\rm DM}_6$), covering the full observation duration of 15,780 s. The red dashed line represents the daily average DM selected by \citet{zhang2026investigating}.}
	\label{dm_time_change1}
\end{figure}

We applied all six DM estimation methods to the 2,874 bursts. Figure~\ref{dm_time_change1} displays the DM time series from each method over the full 15,780~s observation. The six methods produce broadly consistent DM values, clustering around the daily mean of $\sim$529.2~pc~cm$^{-3}$ reported by \citet{zhang2026investigating}. However, individual bursts show visible scatter among the methods, and the degree of scatter varies substantially across the sample.

\begin{figure}
	\centering
	\includegraphics[width=1\columnwidth]{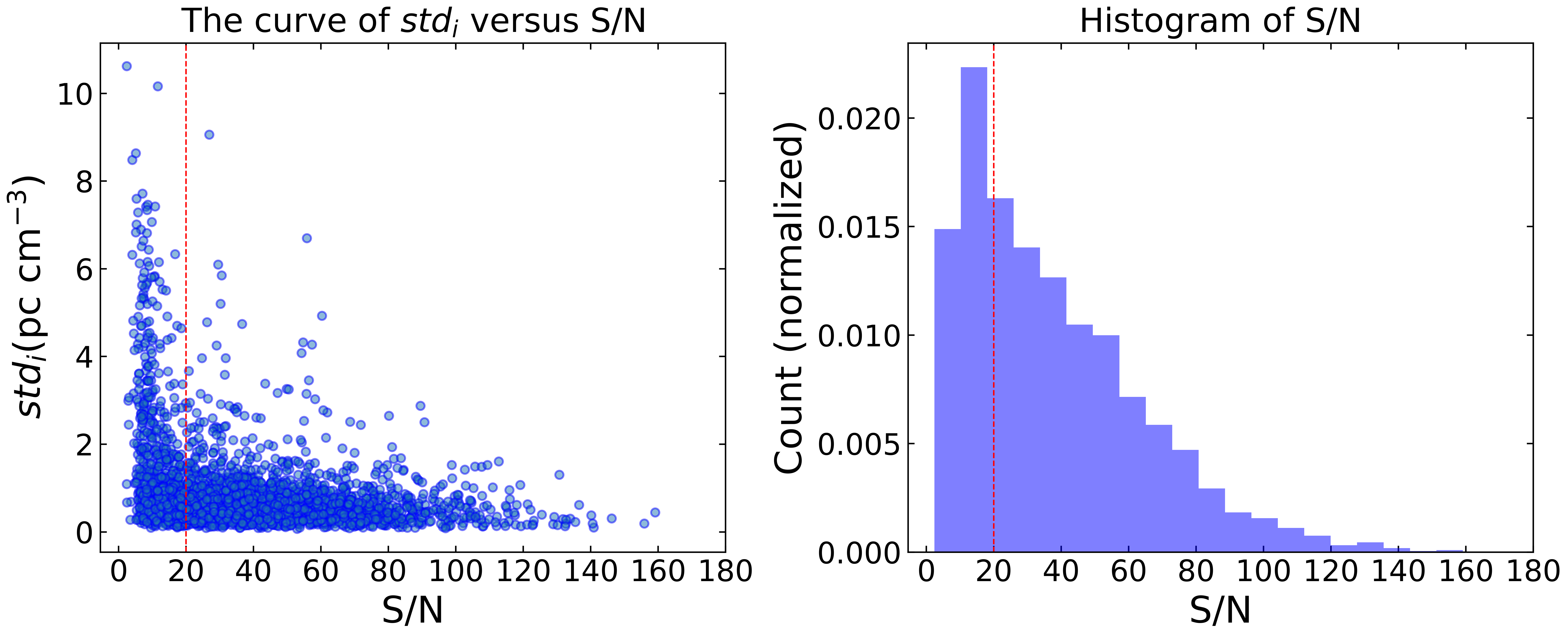}
	\caption{Left panel: the variation of $\mathrm{std}_i$ with S/N for all 2,874 FRB samples, where each point represents one FRB burst. Right panel: the S/N distribution histogram of the 2,874 FRB samples, with the vertical axis showing normalized counts. The red dashed line in both panels marks $\mathrm{S/N} = 20$.}
	\label{s_n20}
\end{figure}

To quantify this scatter, we define std$_i$, the standard deviation of the six DM estimates for each burst (Equation~\eqref{eq:std_i}). Figure~\ref{s_n20}(left panel) shows std$_i$ as a function of S/N. Low-S/N bursts exhibit systematically larger std$_i$, whereas std$_i$ converges at increasing S/N, reflecting the increasing impact of noise-dominated DM optimization for weak signals. The right panel of Figure~\ref{s_n20} shows that most bursts are concentrated at low S/N. To isolate morphology-driven effects from noise-dominated cases, we retain only bursts with S/N $> 20$ \citep{S_N1,S_N2}, yielding 1,944 high-reliability events. This threshold is stricter than the conventional S/N $> 7$ because our data are preprocessed secondary products with incompletely documented RFI mitigation.

\begin{figure}
	\centering
	\includegraphics[width=0.95\columnwidth]{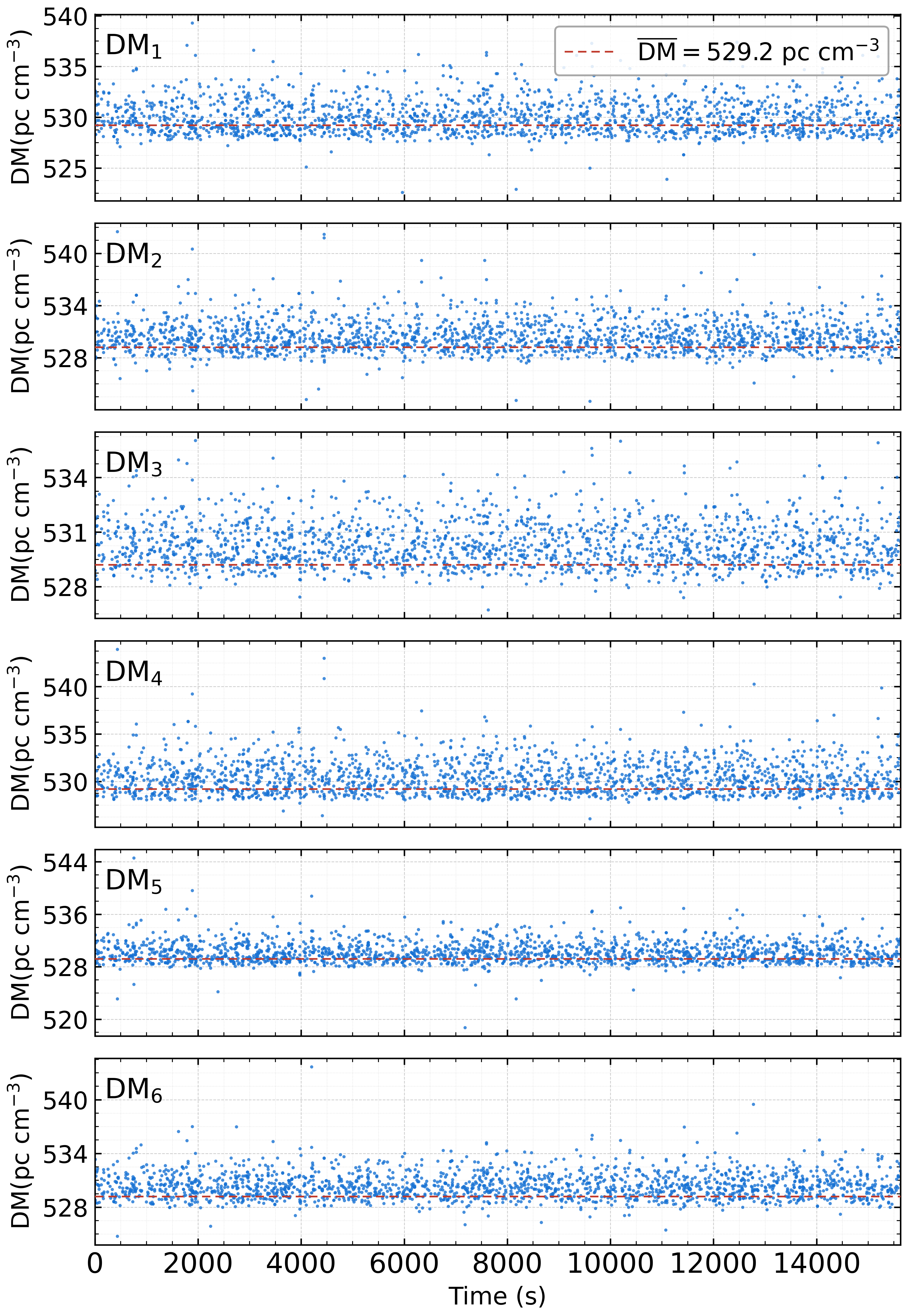}
	\caption{Temporal variations of DM for 1,944 bursts ($\mathrm{S/N} > 20$) of FRB 20240114A observed on March 12, 2024.}
	\label{dm_time_change2}
\end{figure}

Figure~\ref{dm_time_change2} shows the DM time series for the S/N $> 20$ subsample. Even after this cut, individual methods show pronounced fluctuations over the observation.

\begin{figure}
	\centering
	\includegraphics[width=1\columnwidth]{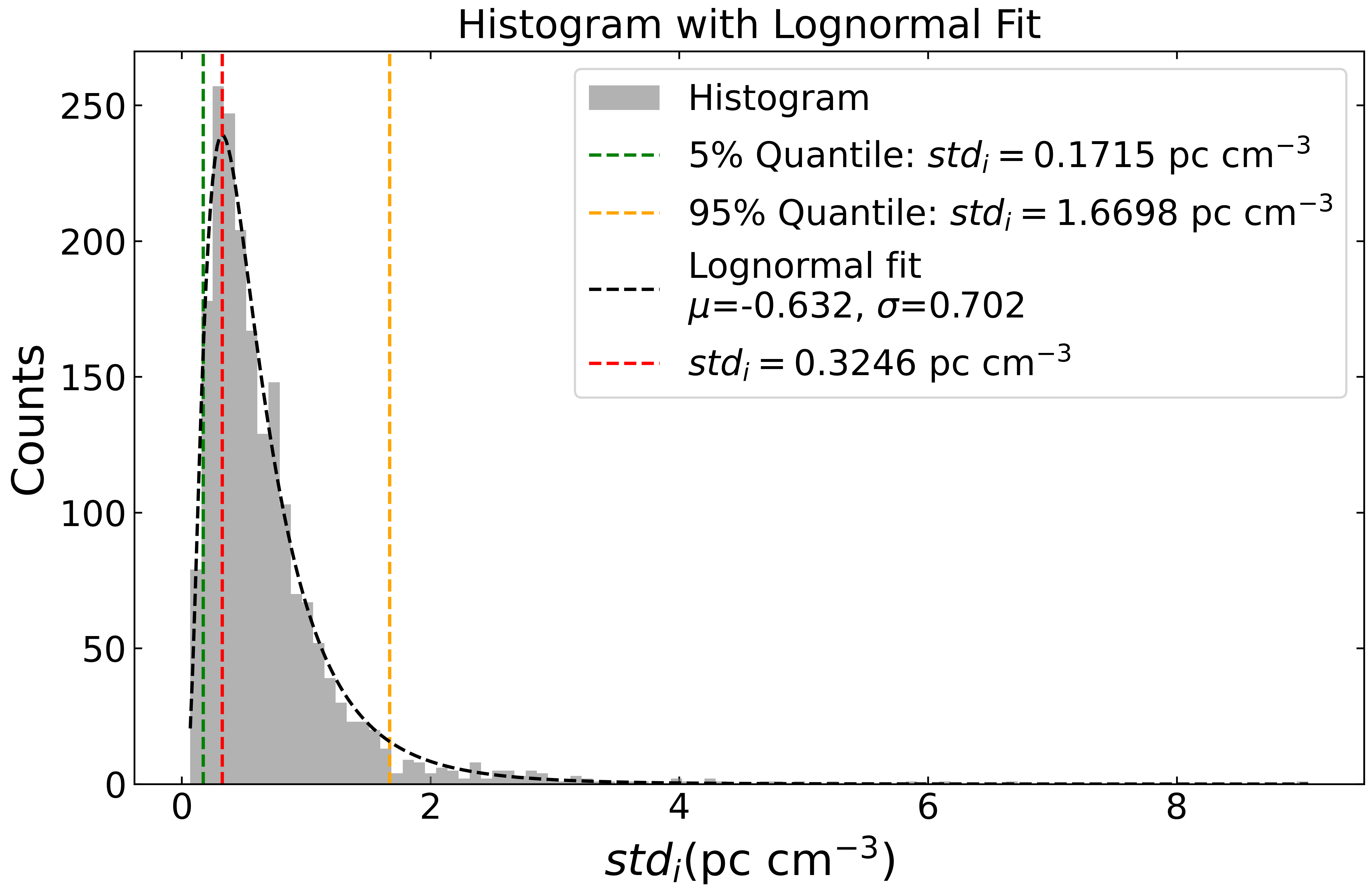}
	\caption{The gray shaded region represents the histogram of $\mathrm{std}$ for 1,944 FRB samples with $\mathrm{S/N} > 20$. The black dashed line represents the log-normal fitting profile with parameters $\mu = -0.632$ and $\sigma = 0.702$. The green and orange dashed lines indicate the 5th and 95th percentiles of the histogram, respectively.}
	\label{histigram}
\end{figure}

Figure~\ref{histigram} shows the std$_i$ distribution for the 1,944 S/N $> 20$ bursts. The distribution is well fit by a log-normal curve ($\mu = -0.632$, $\sigma = 0.702$), with a right-skewed shape peaking near $\sim$0.32~pc~cm$^{-3}$ and a median of $e^{\mu} \approx 0.532$~pc~cm$^{-3}$. Over half the sample thus has inter-method scatter below 0.532~pc~cm$^{-3}$. However, the tail extends to $\sim$8~pc~cm$^{-3}$, indicating anomalously large discrepancies in a small fraction of bursts---likely due to complex morphologies, anomalous RFI, or significant scattering broadening.

To investigate the drivers of high deviation, we select the 98 bursts with std$_i$ at or above the 95th percentile ($1.6698$~pc~cm$^{-3}$) as the high-deviation group, and 98 bursts at or below the 5th percentile ($0.1715$~pc~cm$^{-3}$) as the low-deviation control group.

\begin{figure}
	\centering
	\includegraphics[width=1\columnwidth]{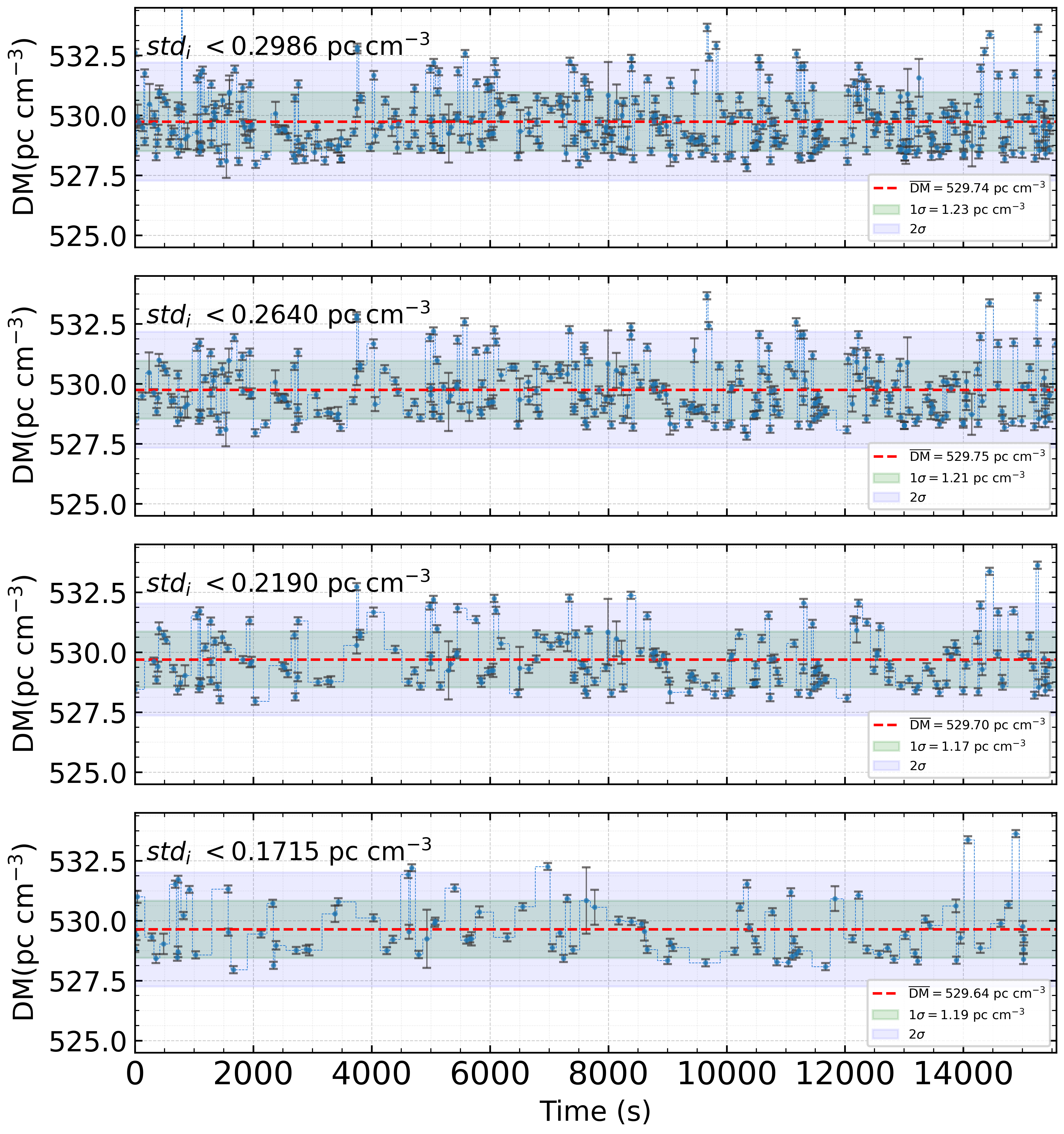}
	\caption{DM distribution over time for bursts from FRB\,20240114A on March 12, 2024 with $\mathrm{S/N} > 20$.
    The four panels show results constrained by inter-method consistency thresholds: $\text{std}_i$ $<$ 0.2986 pc cm$^{-3}$, $\text{std}_i$ $<$ 0.2640 pc cm$^{-3}$, $\text{std}_i$ $<$ 0.2190 pc cm$^{-3}$, and $\text{std}_i$ $<$ 0.1715 pc cm$^{-3}$.
    These thresholds correspond to values below the 20th, 15th, 10th and 5th percentiles in Figure~\ref{histigram}, arranged from top to bottom in that order. The red dashed line denotes the mean DM, and the shaded regions represent the $1\sigma$ and $2\sigma$ standard deviation ranges of the samples. The DM values are derived from the power spectrum method,
with error bars included.}
	\label{dm_stable}
\end{figure}

FRB DM is conventionally expected to remain stable over a $\sim$4-hr observation. To test this, we filter the S/N $> 20$ bursts into four subgroups with progressively stricter std$_i$ constraints and examine DM temporal stability using the power-spectrum measurements. As shown in Figure~\ref{dm_stable}, the DM values exhibit persistent fluctuations spanning $528$--$534$~pc~cm$^{-3}$ even under the strictest constraints (std$_i < 0.1715$~pc~cm$^{-3}$, the 5th percentile).
Crucially, the bursts in this strictest panel are morphologically simple and mutually consistent across all six methods, with per-burst uncertainties (Figure~\ref{dm_stable}) far below the observed $\sim$6~pc~cm$^{-3}$ spread; the residual scatter is therefore neither an algorithmic artefact nor measurement noise. 
The long-term evolution of DM is established \citep{cui, dm_ver1, dm_ver2}, whereas short-term fluctuations of DM remain controversial at present.
We examine its physical origin in Section~\ref{intrinsic_dm}.
This variability demonstrates that apparent DM stability can break down even on short timescales, underscoring the necessity of multi-method cross-validation.

\section{Discussion}
\label{Discussion}


\subsection{Morphology Dependence of DM Deviation}
\begin{figure}
	\centering
	\includegraphics[width=0.95\columnwidth]{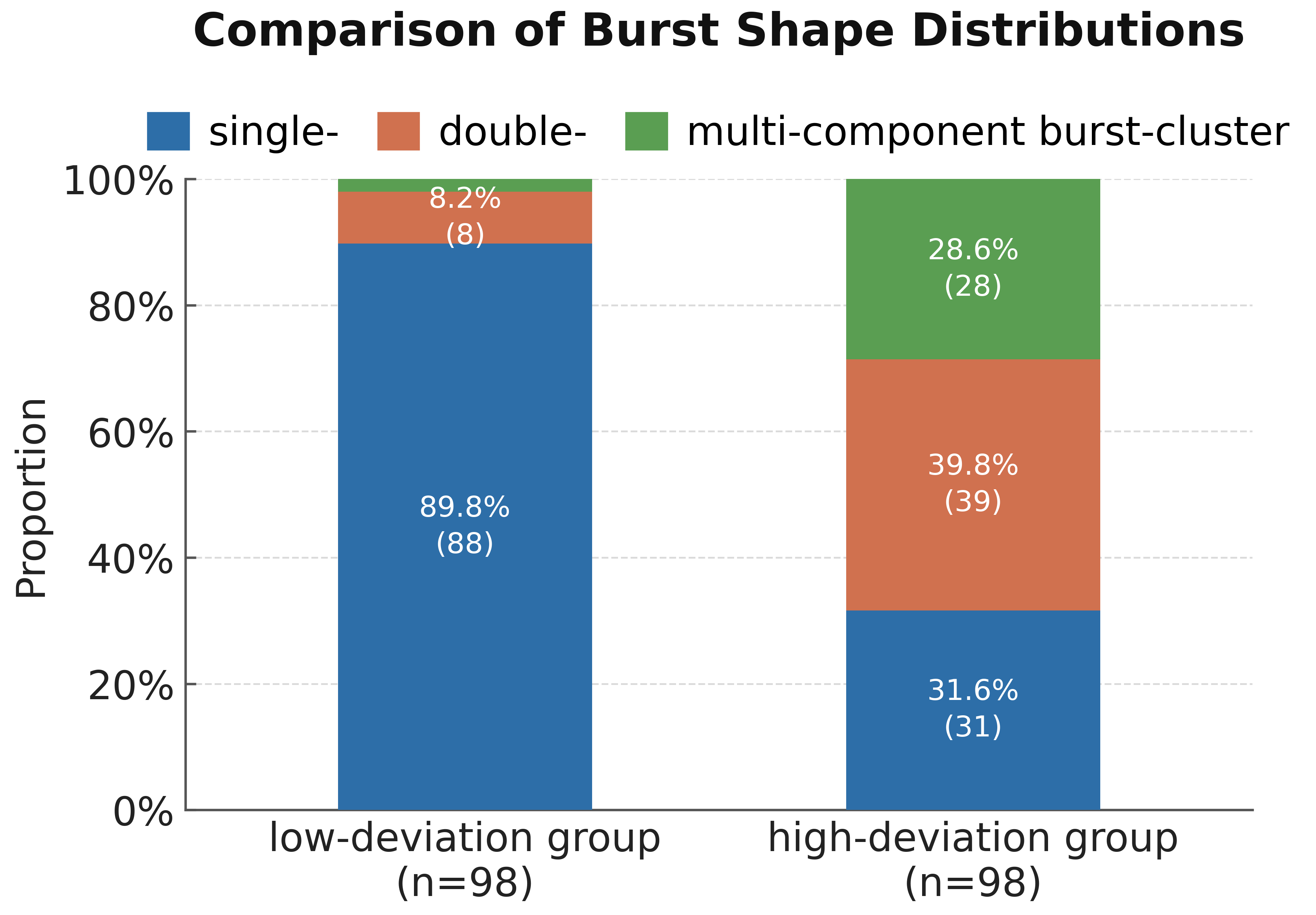}
	\caption{Proportional distribution of FRB burst morphologies in the low-deviation group and the high-deviation group. Blue, red, and green represent single-, double-, multiple-component burst-cluster, respectively. Each group contains 98 bursts.}
	\label{shape_class}
\end{figure}

Figure~\ref{shape_class} compares the morphology distribution between the low- and high-deviation groups. In the low-deviation group, single bursts dominate at $89.8\%$, with double bursts at $8.2\%$ and multiple bursts at $2\%$. Complex morphologies therefore account for only 10.2\% of this group. In the high-deviation group, this reverses: single bursts drop to $31.6\%$, double bursts rise to $39.8\%$, and multiples to $28.6\%$, with complex morphologies totalling $68.4\%$. The proportion of complex morphologies increases by $58.2\%$ between the two groups. This reveals a strong correlation between morphology complexity and inter-method DM discrepancy. For simple single bursts, the dispersion trajectory in the time-frequency plane is unique, producing convergent results across methods. For double and multiple bursts, overlapping dispersion structures cause methods to diverge based on their different objective functions.


\begin{figure}
	\centering
	\includegraphics[width=1\columnwidth]{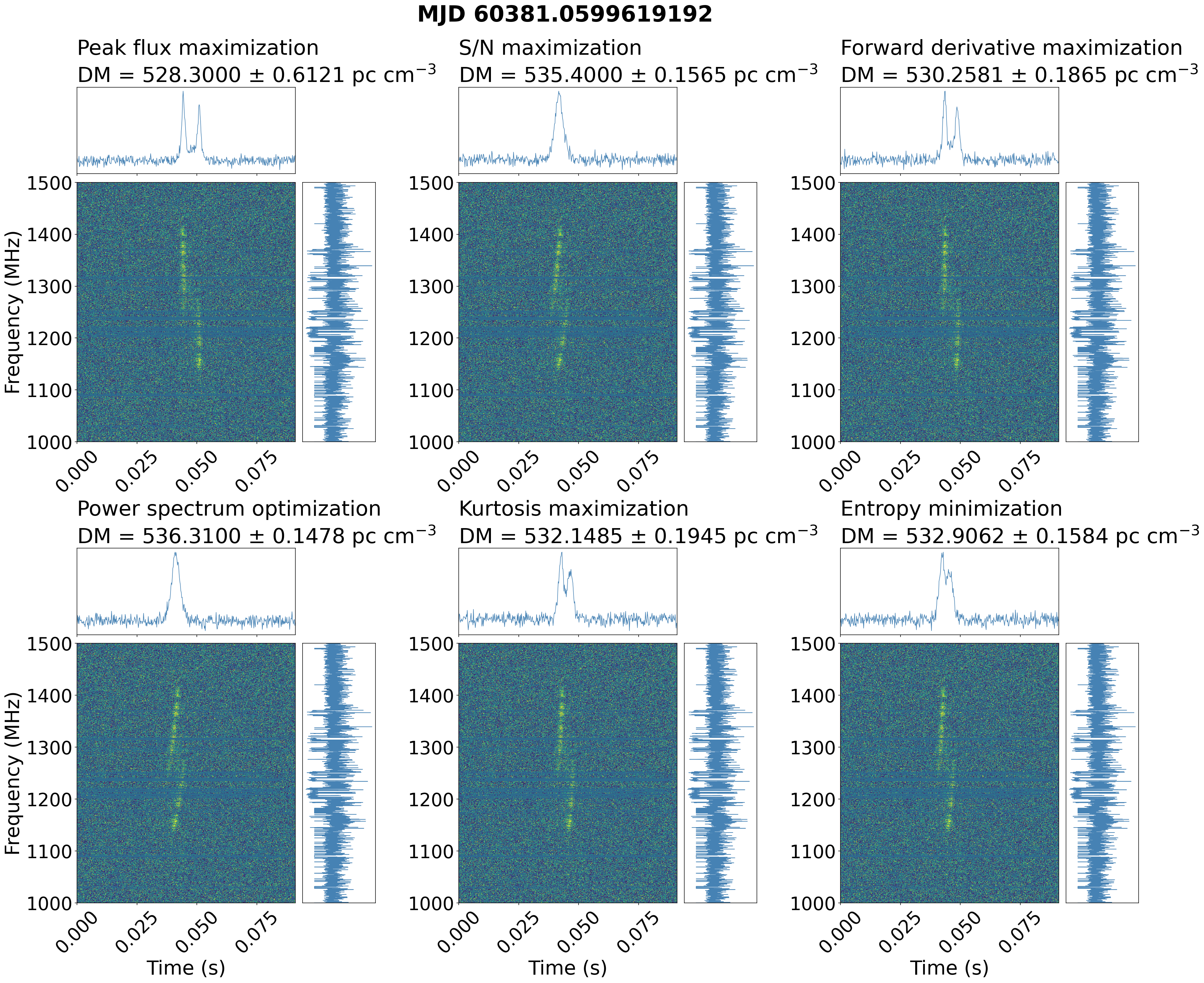}
	\caption{Dynamic spectra of the first burst from the high-deviation group. Panels from left to right display DM measurements obtained by six methods: peak flux maximization, S/N maximization, forward derivative maximization, power spectrum optimization, kurtosis maximization and entropy minimization. The time-integrated and frequency-integrated profiles are included in each sub-panel.}
	\label{high_1}
\end{figure}

\begin{figure}
	\centering
	\includegraphics[width=1\columnwidth]{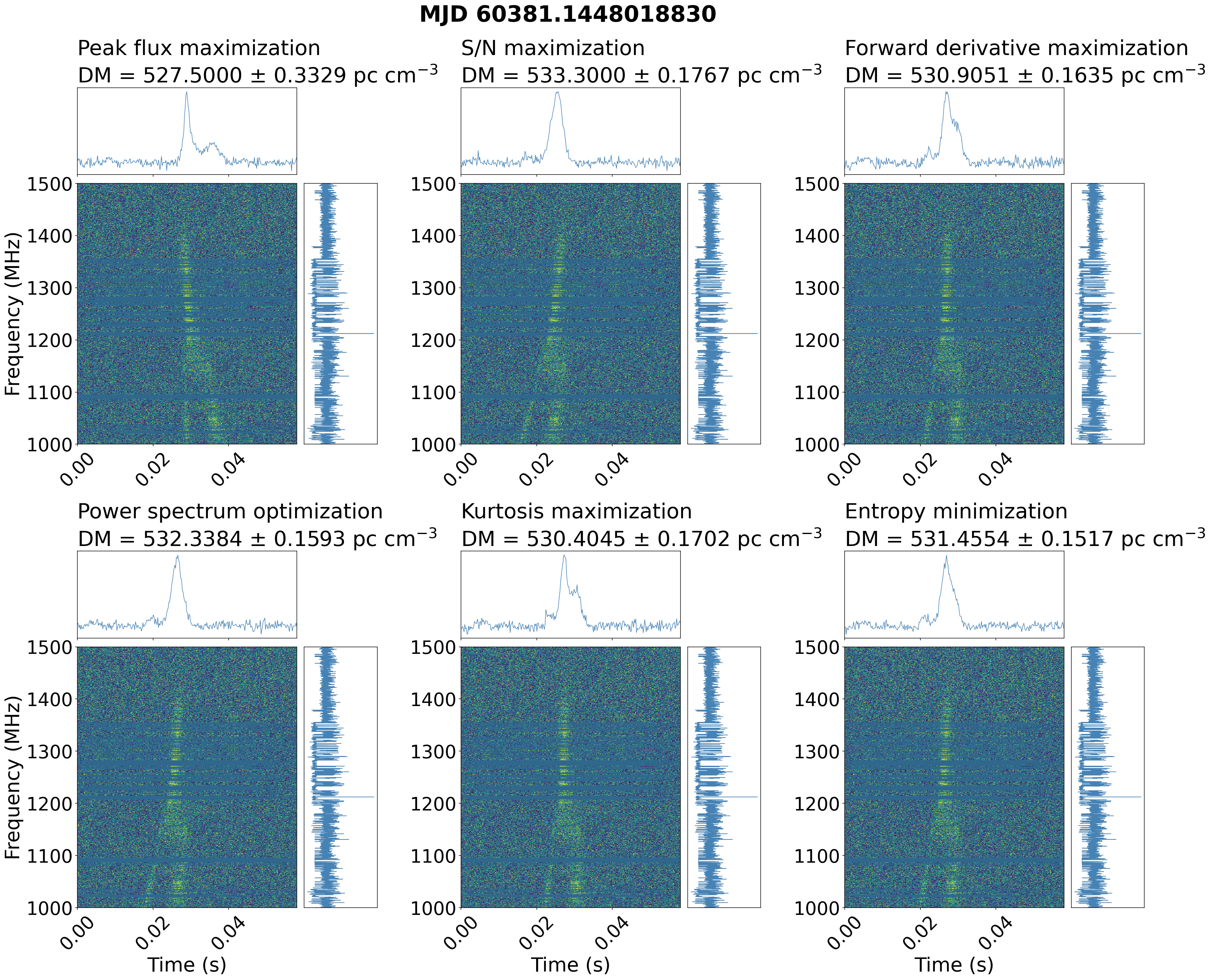}
	\caption{Dynamic spectra of the second burst from the high-deviation group. Details are consistent with those in Figure~\ref {high_1}.}
	\label{high_2}
\end{figure}

\begin{figure}
	\centering
	\includegraphics[width=1\columnwidth]{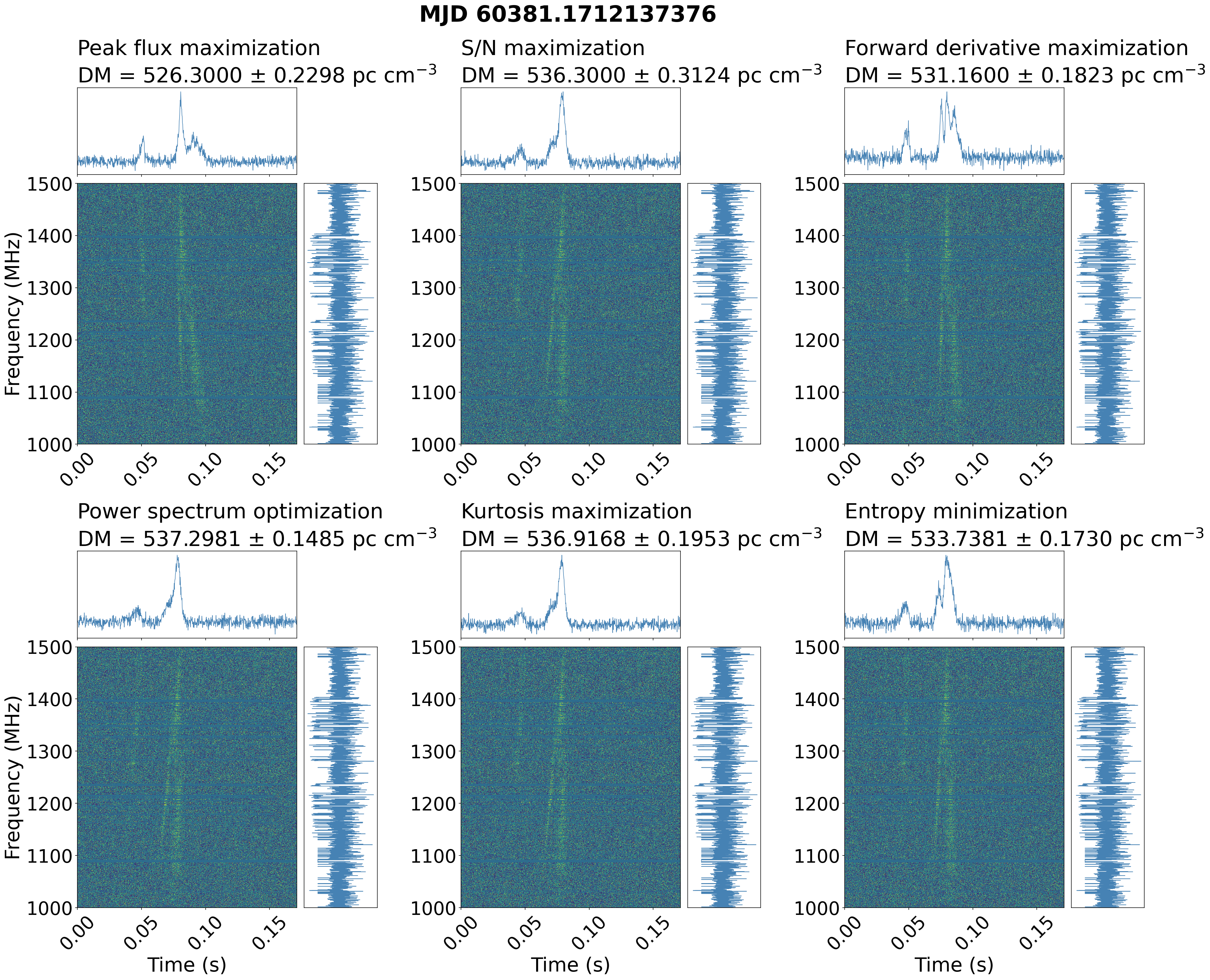}
	\caption{Dynamic spectra of the third burst from the high-deviation group. Details are consistent with those in Figure~\ref {high_1}.}
	\label{high_3}
\end{figure}

\begin{figure}
	\centering
	\includegraphics[width=1\columnwidth]{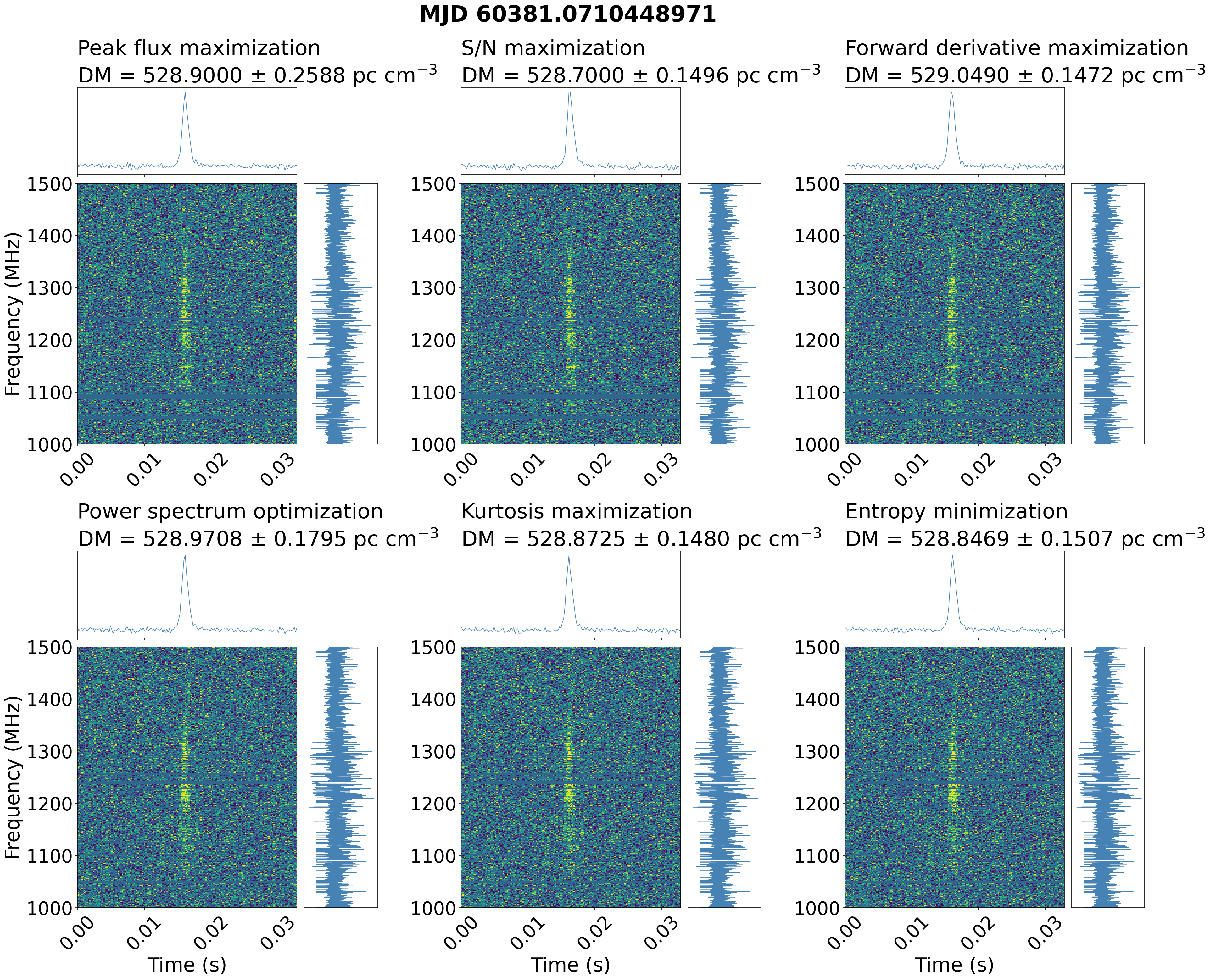}
	\caption{Dynamic spectra of the first burst from the low-deviation group. Details are consistent with those in Figure~\ref {high_1}.}
	\label{low_1}
\end{figure}

\begin{figure}
	\centering
	\includegraphics[width=1\columnwidth]{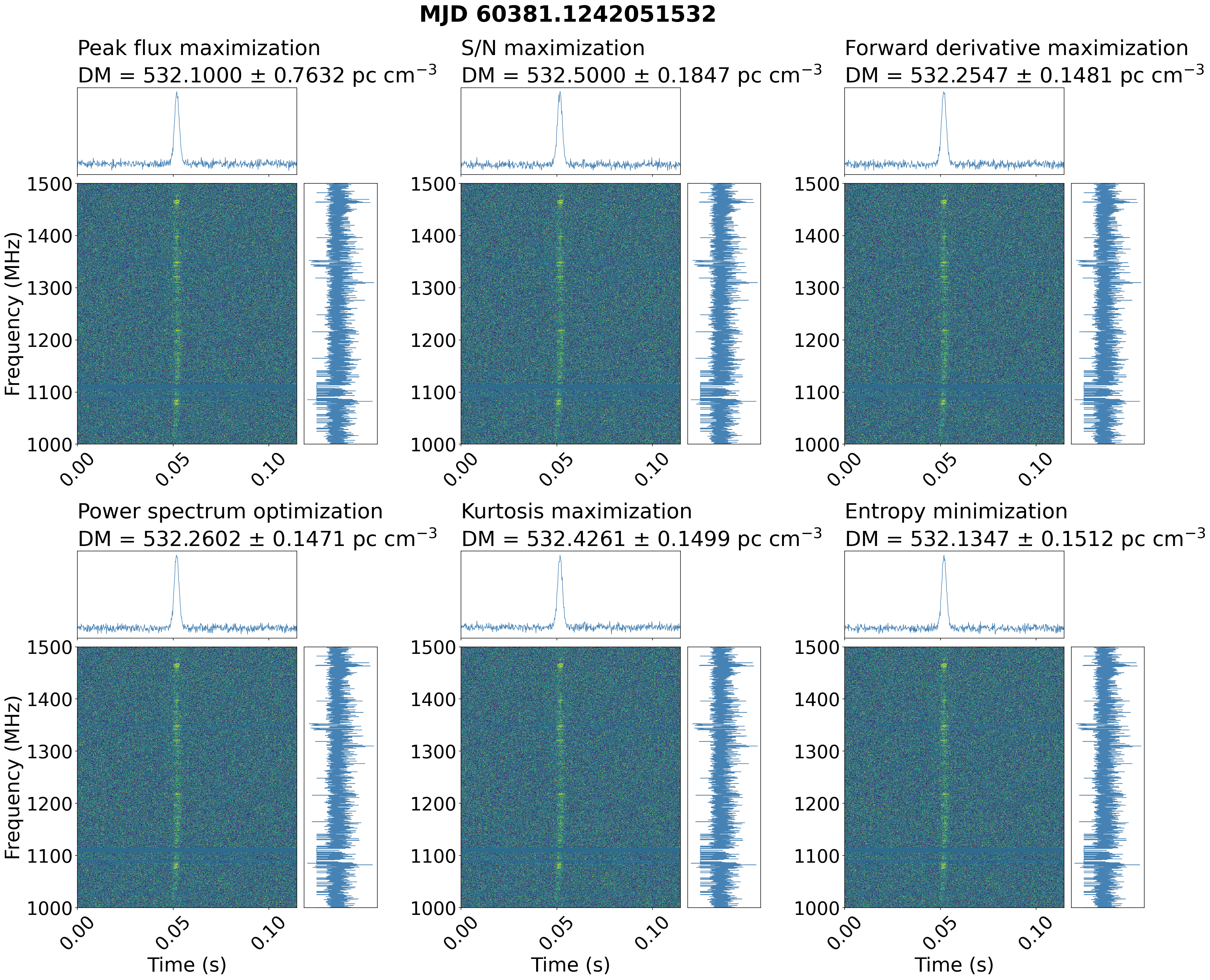}
	\caption{Dynamic spectra of the second burst from the low-deviation group. Details are consistent with those in Figure~\ref {high_1}.}
	\label{low_2}
\end{figure}

Figures~\ref{high_1}--\ref{high_3} show representative high-deviation bursts with drifting substructures. 
For the typical double-component drift structure shown in Figures~\ref{high_1},
The six methods fall into three response categories: Methods 1 and 3 produce the most vertical (linear) de-dispersion trajectories; Methods 2 and 4 give consistently tilted structures; Methods 5 and 6 show intermediate behavior. 
Nevertheless, bursts with intricate substructures in Figures \ref{high_2} and \ref{high_3} do not exhibit such obvious regularities. This indicates that the algorithm yields inconsistent responses to morphologically complex bursts, with complicated overall response behaviors. 
In contrast, for single bursts in the low-deviation group (Figures~\ref{low_1} and \ref{low_2}), all six methods converge to similar DM values. The DM from simple single bursts is thus likely closest to the true value. Morphology complexity is the key factor determining inter-method consistency, for complex bursts, careful method selection is essential.

\begin{figure}
	\centering
	\includegraphics[width=1\columnwidth]{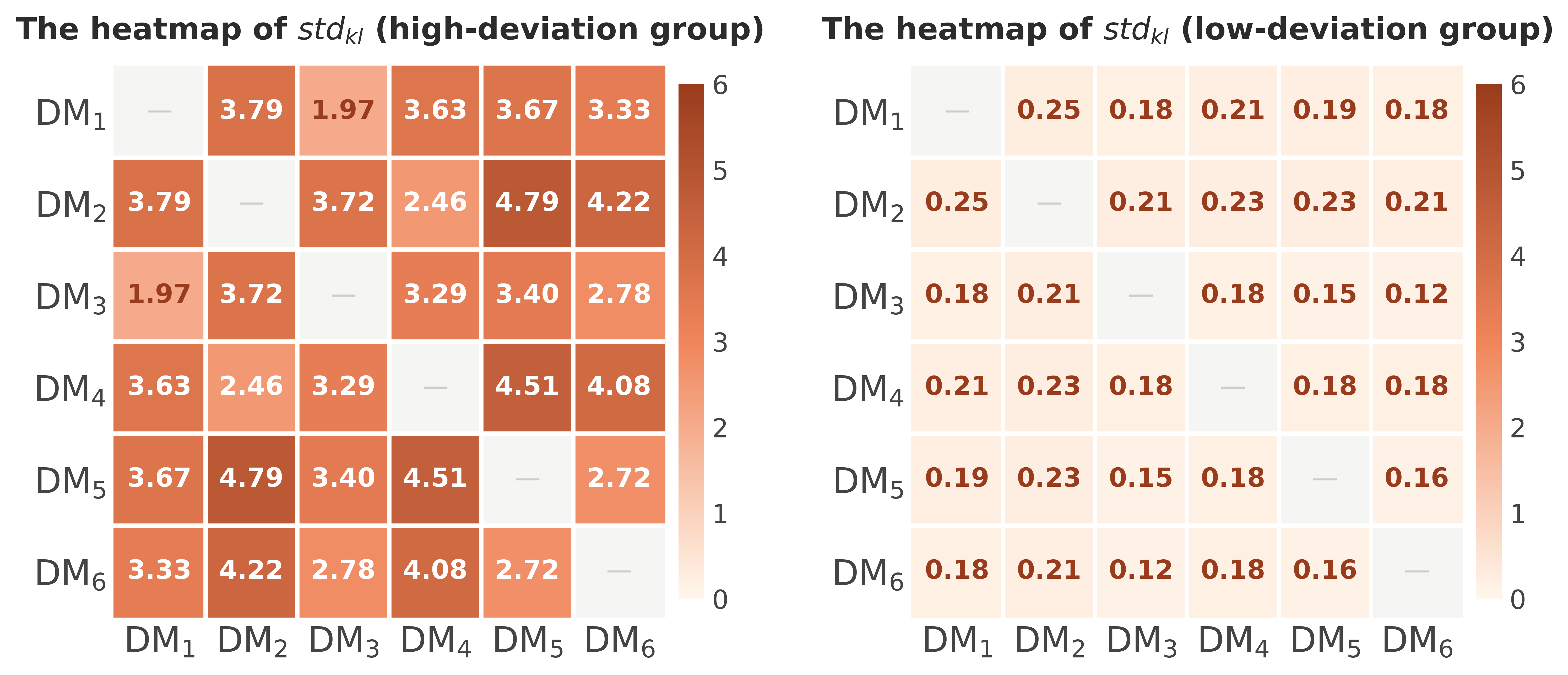}
	\caption{The $\mathrm{std}_{nl}$ heatmaps for the high-deviation group (left panel) 
and the low-deviation group (right panel), where color intensity represents the 
degree of deviation (mapping range: $0$--$6$). Each 
intensity value is derived from Equation~(\ref{eq:std_nl}).}
	\label{heat_1}
\end{figure}

Figure~\ref{heat_1} shows the std$_{kl}$ heatmaps for the high- and low-deviation groups. The high-deviation group spans std$_{kl}$ values of $1.97$--$4.79$ pc cm$^{-3}$, far larger than the $0.12$--$0.25$ pc cm$^{-3}$ in the low-deviation group. In the high-deviation group, three method pairs show relatively small mutual differences: Methods 1--3 (std$_{13} = 1.97$ pc cm$^{-3}$), Methods 2--4 (std$_{24} = 2.46$ pc cm$^{-3}$), and Methods 5--6 (std$_{56} = 2.72$ pc cm$^{-3}$). This ``pairwise proximity'' arises from similar algorithmic responses to multi- and double-component drifting structures, not from intrinsic systematic coupling. In the low-deviation group, these pairwise differences largely vanish (std$_{13}=0.18$, std$_{24}=0.23$, std$_{56}=0.16$ pc cm$^{-3}$). This demonstrates that methods converge when bursts are simple and diverge only when structures are complex.
The methods therefore exhibit both universality for simple bursts and differential sensitivity to complex morphologies, with the latter being the primary origin of the large discrepancies observed in the high-deviation group.

\subsection{Effects of RFI}
\begin{figure}
	\centering
	\includegraphics[width=1\columnwidth]{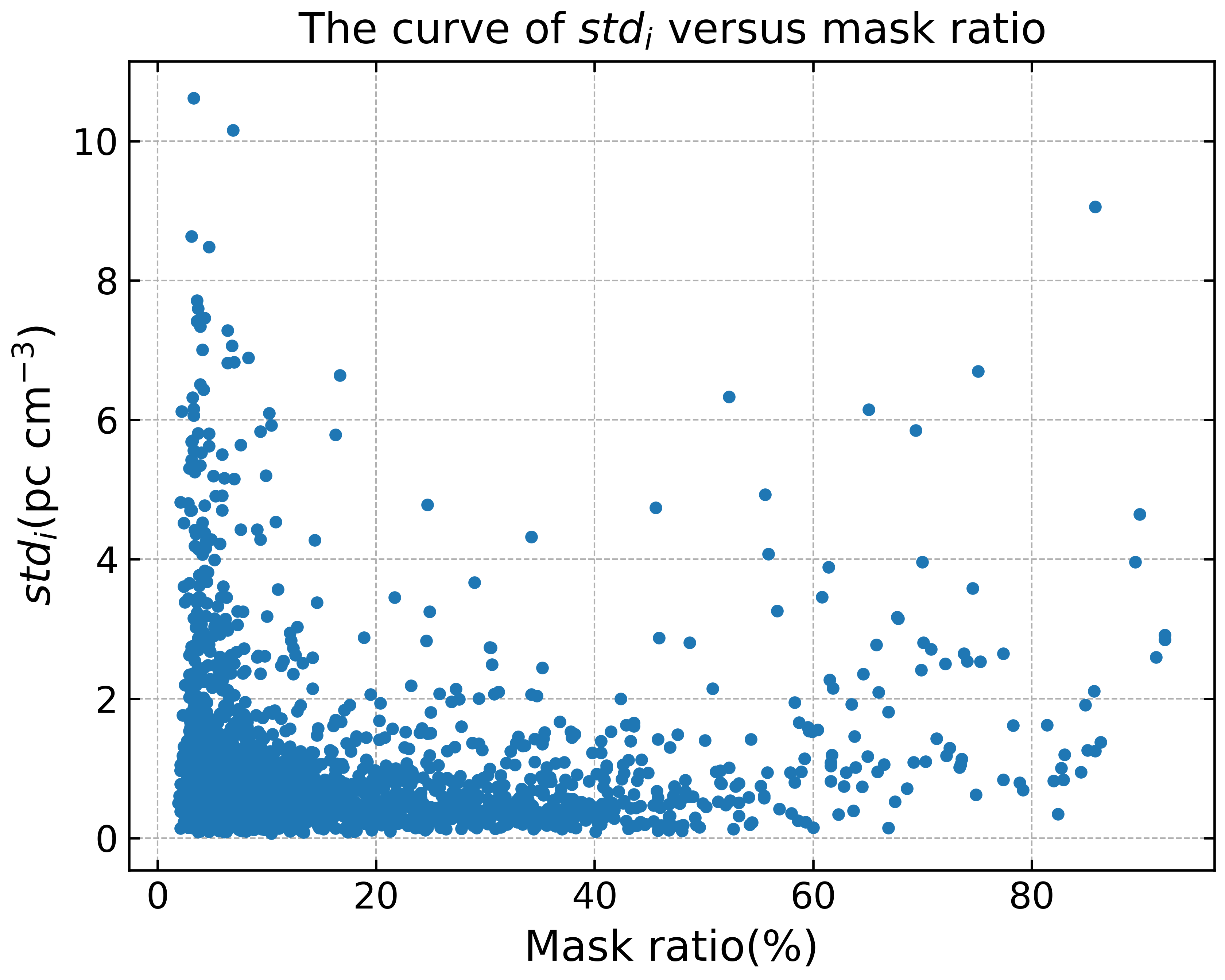}
	\caption{The variation of $\mathrm{std}_i$ with mask ratio (the ratio of frequency channels masked due to RFI) for all 2,874 FRB samples.}
	\label{mask_all}
\end{figure}

Figure~\ref{mask_all} shows std$_i$ as a function of the RFI-masked channel fraction. No systematic trend is observed, indicating that RFI does not significantly affect global DM deviation statistics.

\begin{figure}
	\centering
	\includegraphics[width=1\columnwidth]{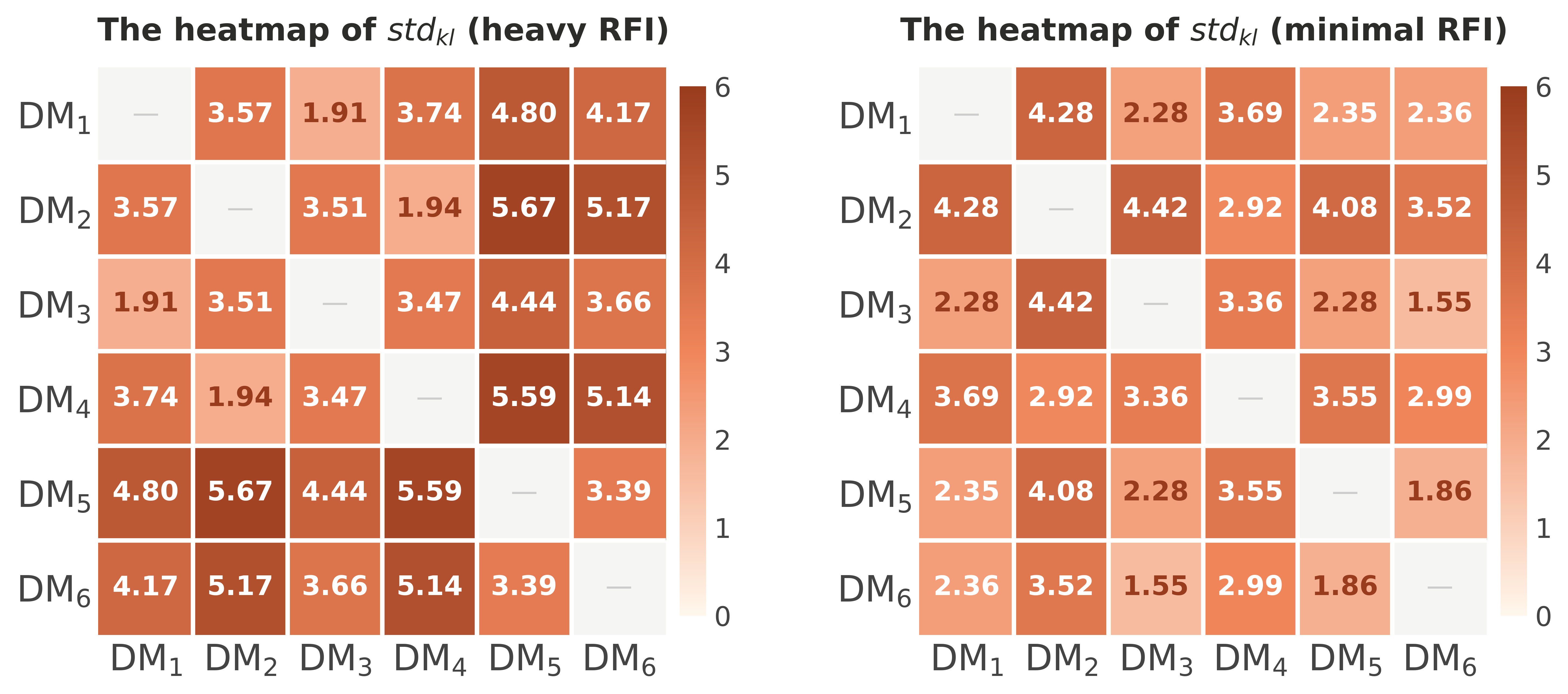}
	\caption{The $\mathrm{std}_{nl}$ heatmaps for the high-deviation group strongly 
affected by RFI (55 samples, left panel) and weakly affected by RFI (43 samples, 
right panel), where color intensity represents the degree of deviation (mapping 
range: $0$--$6$).}
	\label{heat_2}
\end{figure}

\begin{figure}
    \centering
    \includegraphics[width=1\columnwidth]{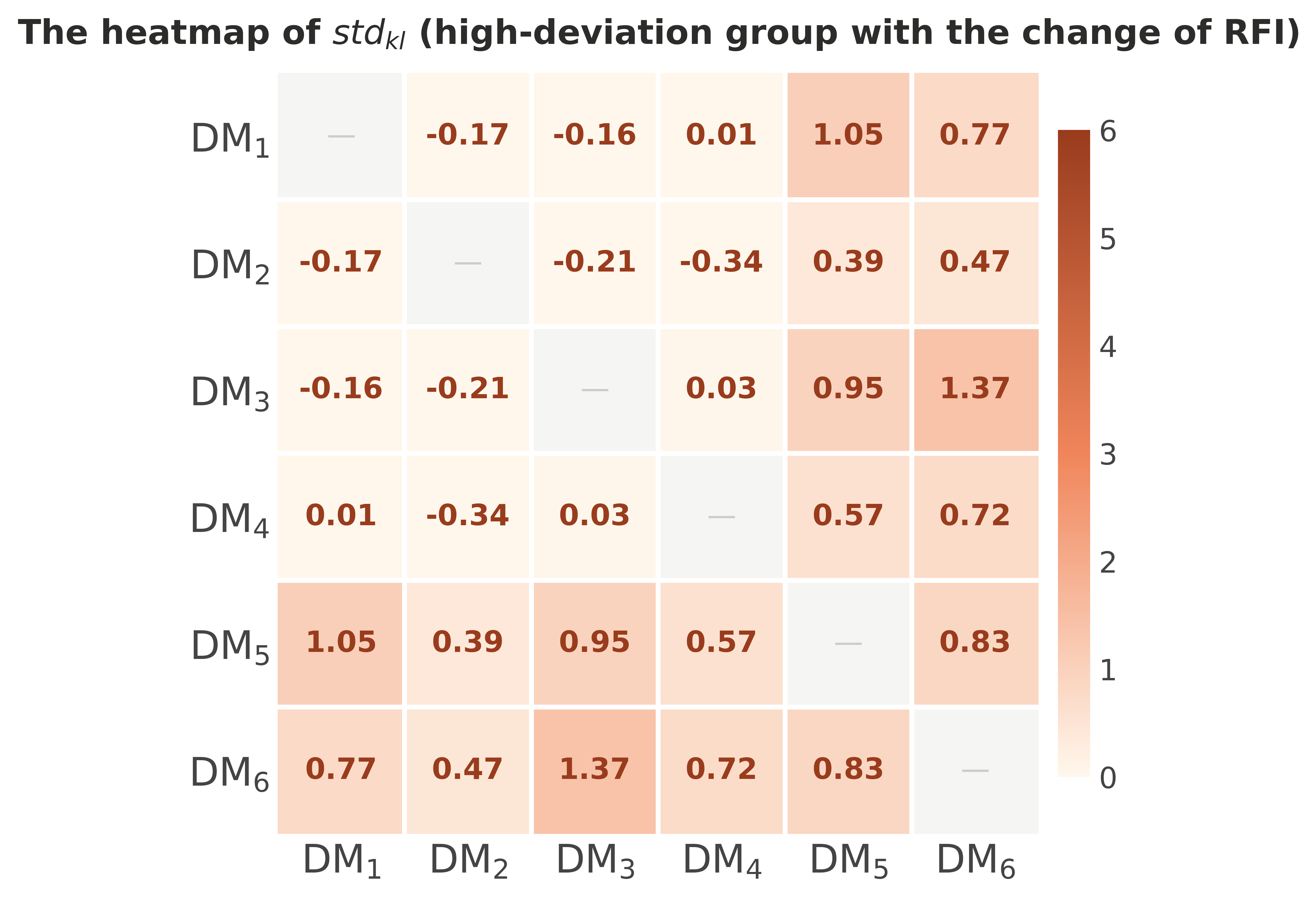}
    \caption{Relative multiplicative change in heatmap intensity between the 
    low-RFI-affected and high-RFI-affected samples shown in Figure~\ref{heat_2}, 
    defined as $\mathrm{std}_{nl} = ({\mathrm{std}_{nl}^{\mathrm{high}} - 
    \mathrm{std}_{nl}^{\mathrm{low}}}) / {\mathrm{std}_{nl}^{\mathrm{low}}}$.}.
    \label{heat_3}
\end{figure}

However, individual methods respond differently. From the high-deviation group, we selected 35 RFI-strong and 63 RFI-weak samples. The std$_{kl}$ difference heatmap (Figure~\ref{heat_2}) and relative change map (Figure~\ref{heat_3}) show that Methods 5 and 6 are notably more sensitive to RFI-induced channel loss. This is likely because density filtering is susceptible to morphology distortion when many frequency channels are masked, especially for complex bursts.

\subsection{Origin of the Short-Timescale Apparent DM Variations}
\label{intrinsic_dm}

\begin{figure}
	\centering
	\includegraphics[width=1\columnwidth]{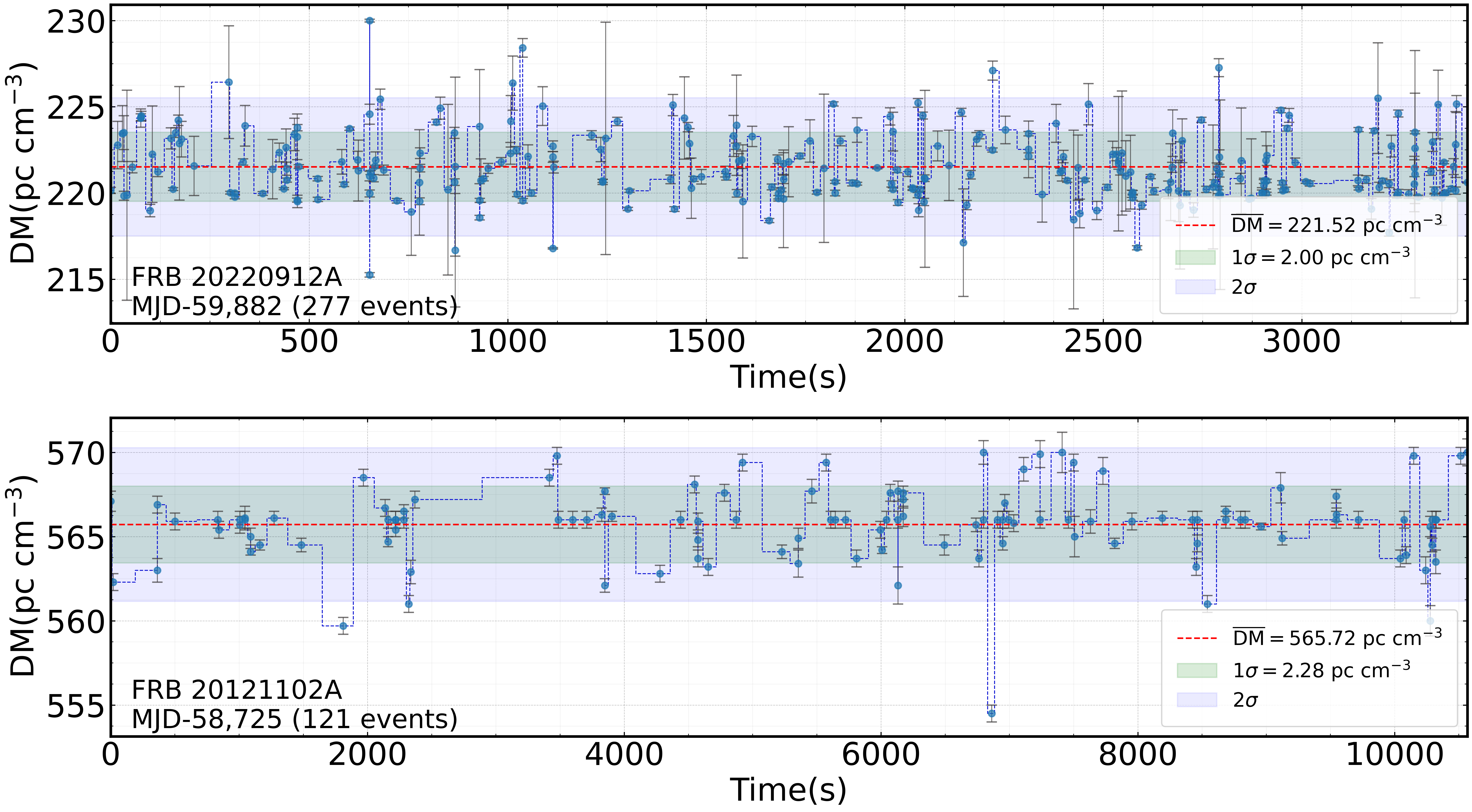}
	\caption{DM distribution over time for FRB\,20220912A (MJD\,59882) and FRB\,20121102A (MJD\,58725), whose observational dates are each selected from the nearly most active day of the respective datasets according to \citet{FRB20220912A} and \citet{FRB20121102A}. The red dashed line denotes the mean DM, and the shaded regions represent the $1\sigma$ and $2\sigma$ standard deviation ranges of the samples.
    }
	\label{other_source}
\end{figure}

DM estimates are derived from morphological analysis rather than from a rigorous analytical formulation\citep{DM_k}. The persistent DM fluctuations in Figure~\ref{dm_stable} are most striking in the strictest-consistency panel
(\({\rm std}_i < 0.1715~{\rm pc~cm^{-3}}\)). In this subset, the bursts are morphologically simple and the six methods give mutually consistent DM estimates. Both the inter-method scatter and the per-burst measurement uncertainties are much smaller than the observed burst-to-burst spread of \(\sim 528\)--\(534~{\rm pc~cm^{-3}}\). The residual variation is therefore unlikely to be produced by algorithmic disagreement or by measurement noise alone. Notably, similar short-timescale apparent DM variations are also detected in other repeating FRBs such as FRB\,20220912A and FRB\,20121102A (see Figure~\ref{other_source}). The remaining question is whether it reflects a genuine change in the line-of-sight electron column, or an apparent DM induced by burst-intrinsic or local propagation effects.

Using the standard dispersion relation \citep{lorimer2005handbook}, an apparent DM change \(\Delta{\rm DM}\) corresponds to a relative arrival-time offset between the lower and upper band edges of
\begin{equation}
\Delta t_{\rm band}
=
K\,\Delta{\rm DM}\left(\nu_{\rm low}^{-2}-\nu_{\rm high}^{-2}\right)
\simeq
2.3~{\rm ms}
\left(
\frac{\Delta{\rm DM}}{1~{\rm pc~cm^{-3}}}
\right),
\label{eq:dm_time_offset}
\end{equation}
for the 1.0--1.5 GHz FAST band, where \(K\) is the dispersion constant defined in Equation~(\ref{eq:K}). Thus, an apparent DM excursion of \(1\)--\(6~{\rm pc~cm^{-3}}\) corresponds to only a few to \(\sim 10\) ms of relative timing shift across the band, comparable to the intrinsic widths, unresolved substructures, and frequency drifts commonly seen in repeating FRB bursts.

A literal propagation interpretation would require the electron column to vary by
\begin{equation}
\Delta N_e
=
\Delta{\rm DM}\times 3.086\times10^{18}~{\rm cm^{-2}}
\simeq
3.1\times10^{18}
\left(
\frac{\Delta{\rm DM}}{1~{\rm pc~cm^{-3}}}
\right)
{\rm cm^{-2}} .
\label{eq:delta_ne_column}
\end{equation}
Adjacent bursts in the strict subsample are separated by only seconds to minutes, yet their apparent DMs can differ by \(\gtrsim 1~{\rm pc~cm^{-3}}\). If the variation were produced by plasma of density \(n_e\) moving across the effective line of sight at speed \(v\), then \(\Delta N_e \simeq n_e v\Delta t\), giving
\begin{equation}
n_e
\simeq
\frac{\Delta N_e}{v\Delta t}
\gtrsim
1.0\times10^{6}
\left(
\frac{\Delta{\rm DM}}{1~{\rm pc~cm^{-3}}}
\right)
\left(
\frac{v}{c}
\right)^{-1}
\left(
\frac{\Delta t}{100~{\rm s}}
\right)^{-1}
{\rm cm^{-3}} .
\label{eq:density_requirement}
\end{equation}
This is already a demanding requirement even in the limiting case \(v=c\). For a more typical outflow or shocked-plasma velocity of \(v\sim10^3~{\rm km~s^{-1}}\), the required density increases to \(\gtrsim 3\times10^8~{\rm cm^{-3}}\). Such a dense, fast-moving screen is difficult to accommodate in the local environment of an FRB. Moreover, a medium variable enough to produce minute-scale pc~cm$^{-3}$ jumps would be difficult to reconcile with the stable daily-mean DM of \(\sim529.2~{\rm pc~cm^{-3}}\) reported for this session \citep{zhang2026investigating}. By comparison, secular DM evolution reported in repeating FRBs is usually discussed on timescales of days or longer, far slower than required here \citep{NiuCH2026}. In addition, a rapidly varying plasma screen would likely produce correlated changes in scattering, absorption, RM, depolarization, or spectral modulation. Local plasma associated with magnetar activity, such as flare ejecta, may contribute to longer-term or isolated DM/RM excursions \citep{XiaoD2025}, but it would more naturally produce a smooth trend or a single event-like deviation rather than the repeated, non-monotonic burst-to-burst jumps seen in Figure~\ref{dm_stable}.

A more natural interpretation is that the measured quantity is an apparent DM, rather than a pure propagation DM. 
As indicated in \cite{51.5}, the ``real DM'' might be able to be derived from the high time resolution microshots. As shown in Figure 2 of  \cite{51.5}, the ``real DM'' keeps unchanged for a long period up to tens of day, while the ``apparent DM'' is highly diverse. This is quick similar in the current work.
If the emission time depends on observing frequency, for instance, the downward-drifting ``sad-trombone'' effect \citep{51}, the observed arrival time can be written schematically as
\begin{equation}
t_{\rm obs}(\nu)
=
t_0
+
K\,\,{\rm DM}_{\rm true}\nu^{-2}
+
t_{\rm int}(\nu),
\label{eq:intrinsic_delay}
\end{equation}
where \(t_{\rm int}(\nu)\) represents the intrinsic chromatic emission-time structure of the burst. Over a limited observing band, part of \(t_{\rm int}(\nu)\) can be absorbed by a best-fit \(\nu^{-2}\) dispersive delay, yielding
\begin{equation}
{\rm DM}_{\rm app}
=
{\rm DM}_{\rm true}
+
\Delta{\rm DM}_{\rm int}.
\label{eq:apparent_dm}
\end{equation}
In this picture, different bursts can have different apparent DMs even if the true propagation DM remains nearly constant. Since \(\Delta{\rm DM}=1~{\rm pc~cm^{-3}}\) corresponds to only \(\sim2.3\) ms across the FAST band, modest burst-to-burst changes in intrinsic frequency--time structure can mimic the observed \(1\)--\(6~{\rm pc~cm^{-3}}\) apparent DM offsets.

Structured local propagation may also contribute. Plasma lensing or multi-path propagation can introduce frequency-dependent delays and cause different bursts, or different frequency components of the same burst, to sample different effective ray paths. Such effects would again appear as apparent DM variations rather than as a rapid change in the total electron column along a fixed line of sight. A useful distinction is that intrinsic chromatic emission should mainly correlate with burst morphology, drift rate, or spectral occupancy, whereas local propagation effects should more often correlate with scattering time, RM, polarization behavior, or lensing-like spectral structure.

We therefore interpret the rapid DM fluctuations in Figure~\ref{dm_stable} primarily as apparent DM variations. A genuine change in the line-of-sight electron column is difficult to exclude completely, but it would require a compact, dense, and rapidly evolving plasma structure close to the source. The observed non-monotonic behavior is more naturally explained by burst-dependent frequency--time structure, possibly with a subdominant contribution from structured local propagation. Future broadband observations with simultaneous measurements of DM, RM, scattering time, polarization, and spectral morphology will be crucial for breaking the degeneracy of these possibilities.

\section{Conclusion}
\label{Conclusion}

We have presented a systematic comparison of six DM estimation methods using 2,874 bursts from FRB\,20240114A, currently the most active known repeating FRB, observed by FAST during a single 4.4 hr session. This large and homogeneous sample, obtained over a short timescale during which the propagation environment is expected to remain nearly stable, provides an ideal benchmark for isolating algorithm-dependent effects in DM measurements from genuine astrophysical variations. The kurtosis-maximization and entropy-minimization methods based on density filtering were introduced for the first time, extending DM measurements from traditional signal- and structure-optimization toward statistical characterization in SDS.
%
Even after strict selection for low-scatter bursts, FRB\,20240114A exhibits notable apparent DM fluctuations of $528$--$534~\mathrm{pc\ cm^{-3}}$ over 15,780~s. 

This puts us in a dilemma. If the DM is correct, what causes such big DM variation in a so short timescale is quite challenging.  While if we set the real DM value as unchanged, what caused the apparent extra DM is also puzzling. 
Currently, the later explanation is more probable. 
As shown in Section~\ref{intrinsic_dm}, these second-to-minute variations cannot be produced by genuine changes in the line-of-sight electron column nor by measurement uncertainty, and most plausibly reflect a frequency-dependent emission-time structure intrinsic to the bursts that mimics a dispersive delay. If confirmed, apparent DM variations may serve as a new probe of the FRB emission mechanism.

The consistency of DM measurements depends strongly on S/N, burst morphology, and RFI conditions. Low-S/N bursts show substantially larger inter-method deviations due to noise-dominated uncertainties. Single-component bursts produce highly consistent DM values, while complex multi-component bursts with drifting substructures lead to significant scattering. Pairwise deviation analysis reveals that different methods exhibit distinct responses to complex time-frequency structures, indicating that discrepancies are driven by burst morphology rather than intrinsic algorithmic biases. RFI primarily affects density-filtering methods through morphology distortion from channel masking, though it does not significantly alter the global statistical behaviour.

Overall, the reliability of FRB DM measurements is closely linked to burst morphology and method-dependent sensitivity to complex structures. Multi-method cross-validation and explicit treatment of morphology-dependent systematic uncertainties will therefore be essential for future high-precision FRB studies.


\section*{DATA AVAILABILITY}
The dataset underlying this work used is available at the Science Data Bank \url{https://doi.org/10.57760/sciencedb.Fastro.00033} for the day on 12 March, 2024.
\label{code}
The code used in this work is available at: \url{https://github.com/loganlun/DM_cross.git}.

\section*{Acknowledgments}
This work made use of the data from FAST FRB Key Science Project. This work is supported by the National Natural Science Foundation of China (grant Nos.\ 12503058, 12588202, and 12041306), the Postdoctoral Fellowship Program of CPSF (grant No.\ GZC20252100), the ACAMAR Postdoctoral Fellowship, the China Postdoctoral Science Foundation (grant No.\ 2025M773201), and the Jiangsu Funding Program for Excellent Postdoctoral Talent. P.W. acknowledges support from the CAS Youth Interdisciplinary Team, the Youth Innovation Promotion Association CAS (id. 2021055), and the Cultivation Project for FAST Scientific Payoff and Research Achievement of CAMS-CAS.

\bibliography{sample701}{}
\bibliographystyle{aasjournalv7}



\end{document}